\documentstyle[twoside,fleqn,espcrc2]{article}

\input{psfig}
\newcommand\be{\begin{equation}}
\newcommand\ee{\end{equation}}
\newcommand\bea{\begin{eqnarray}}
\newcommand\eea{\end{eqnarray}}
\newcommand\half{{\textstyle{1\over2}}}

\def\d{\partial}

\title{\Large\bf Instantons in the Maximally Abelian Gauge\thanks{This work
                 is supported in part by funds provided by D.O.E. 
                 under contract \#DE-FG02-91ER400688,Task A. Computational
                 work in support of this research was performed
                 at the Theoretical Physics Computing Facility at 
                 Brown University.
                 Talk presented by K.N. Orginos.}}

\author{R.C. Brower,\address{Department of Physics,
        Boston University, 590 Commonwealth Ave.,
        Boston, MA 02215, USA}
        K.N. Orginos and  C-I Tan\address{Department of Physics,
        Brown University, Providence, RI 02912, USA}}

\begin{document}

\begin{abstract} 

  We investigate the Maximally Abelian (MA) Projection for a
single $SU(2)$ instanton in continuum gauge theory.  We find that
there is a class of solutions to the differential MA gauge condition
with circular monopole loops of radius $R$ centered on the instanton
of width  $\rho$.  However, the MA gauge fixing functional $G$ decreases
monotonically as $R/\rho \rightarrow 0$. Its global minimum  is the 
instanton in the singular gauge. We point out that interactions
with nearby anti-instantons are likely to excite these monopole loops. 

\end{abstract}

\maketitle

\section{Introduction}
 The Abelian projection reduces a non-Abelian gauge theory to a theory
of monopoles and Abelian gauge fields interacting with charged gluons.
It has been claimed both by numerical
simulations\cite{Markum,Schierh,Tanaka} and analytical
calculations\cite{Tanaka,Gubar,bto} that there is a correlation
between magnetic currents of the Abelian projected fields and
instantons of the non-Abelian theory. Here we report on a thorough
analytical and numerical study of the MA Abelian projection of the
instanton.

 In our investigation we use the widely accepted definition for the
Maximally Abelian (MA) gauge\cite{KSW}, the minimization of the functional
\begin{equation}
G = {1 \over 4} \int d^4x \{ A^1_{\mu}(x)^2+A^2_{\mu}(x)^2\},
\label{MA_func}
\end{equation}
or in differential form $(\partial_\mu \pm i e A^3_\mu(x))A^\pm_\mu(x)=0$. 

 There are three key points to our results: (a) Instantons of width
$\rho$ do contain magnetic monopole loops of radius $R$. (b) The
functional $G$ decreases monotonically as $(R/\rho)^4 \log(R/\rho)$ as
$R/\rho \rightarrow 0$, thus small monopole loops are favored.  (c)
Interactions with nearby anti-instantons stabilize monopole loop
formation.

\section{Instanton in the MA gauge }

 We would like to find the gauge transformation $\Omega$ that rotates an
instanton to the MA gauge. $\Omega$  satisfies 
\begin{equation}
D^2_\mu(A)\vec\Phi+\sigma \vec \Phi=0, \>\> |\vec\Phi|={\bf 1},
\label{diff_cond}
\end{equation}
where 
$\vec\Phi(x)\cdot \vec \tau \equiv \Phi(x)= \Omega \tau_3 \Omega^{\dagger}$,
and $\sigma$ is a Lagrange multiplier.
 The $SU(2)$ instanton field in the singular gauge is given by
\begin{equation}
A^{s}_\mu = -\tau^\alpha \bar\eta^\alpha_{\mu\nu}x_\nu 
        \frac{1}{x^2}\frac{\rho^2}{x^2+\rho^2}.
\label{inst_sing}
\end{equation} 
 It can be readily shown that this configuration satisfies the MA gauge
condition.
The functional $G$ is finite and takes the value $G=4\pi^2\rho^2$.
Similarly the instanton in the non-singular gauge,
\begin{equation}
A^{ns}_\mu = -\tau^\alpha \eta^\alpha_{\mu\nu}x_\nu \frac{1}{x^2+\rho^2},
\label{inst_non-sing}
\end{equation}
also satisfies the differential MA condition, but $G$ diverges.
We observe that, in the singular gauge,  the largest contribution to $G$ comes
from the region near the instanton center and it is well-behaved 
at infinity. On the other hand, for
the non-singular gauge,  the contribution to $G$ from the origin
is suppressed and the divergence 
comes from the infinity. It is possible to consider  an 
intermediate gauge where the gauge potential
approaches  the behavior  of  the singular gauge at infinity  
and the behavior  of the non-singular gauge at the origin.
 This is our key idea in searching for solutions
to (\ref{diff_cond}). Since the instanton in the singular gauge already
yields a finite value for 
$G$, we will use it  as the starting point of our investigation. 

We parameterize $\vec\Phi$, which is 
an iso-vector  of unit length, by  its spherical  coordinates: the
polar angle $\beta$ and the azimuthal angle $\alpha$. In this parameterization,
equation (\ref{diff_cond}) becomes
\begin{equation}
\partial^2_\mu \beta-\half \sin(2\beta) (\partial_\mu\alpha)^2=
2\sin\beta \partial_\mu\alpha \bar{A}^3_\mu
\label{beta_eq}
\end{equation}
\begin{equation}
\partial^2_\mu \alpha+2\cot \beta \partial_\mu\alpha\d_\mu\beta
=-2 \partial_\mu\beta \bar{A}^3_\mu
\label{alpha_eq}
\end{equation} 
where $\bar{A}^3_\mu = \vec A_{\mu}\cdot\vec\Phi $.

One solution to (\ref{diff_cond}) corresponds to that discussed  
by Chernodub and Gubarev\cite{Gubar}: $\beta=\vartheta$ and 
$\alpha=\varphi $,
where $\vartheta$ and $\varphi$ are the polar and azimuthal angles 
for the spatial three-vector
$\vec x$. 
 This static solution leads to a divergent $G$ and certainly is not
preferred in the single instanton case.

 Another obvious solution to  (\ref{diff_cond})
corresponds to the one that rotates the singular gauge to the
non-singular gauge.
In our parameterization that solution is $\beta = 2\theta$ and 
$\alpha = \phi - \psi $, where $\theta \equiv \tan^{-1} (u/v)$, 
$\phi \equiv  \tan^{-1} (y/x) $,
$\psi \equiv \tan^{-1} (t/z) $ and $v^2\equiv t^2+z^2$, $u^2\equiv x^2+y^2$. 

 If one assumes $\alpha = \phi - \psi $ and $\beta = \beta(u,v)$, Eq.
(\ref{alpha_eq}) is automatically satisfied.  This ansatz allows
solutions with a monopole loop in the 3-4 and/or 1-2 planes. 
All other allowed orientations of the monopole
loop can be generated by one of the chiral SU(2)/U(1) cosets of the
Lorentz group.

Having in mind  a solution with $A_\mu(x)$ behaving like  the 
non-singular gauge at the origin, and like  the singular gauge
at infinity, we can consider  a variational ansatz for $\beta$,
\begin{equation}
\beta_0(x,\theta)\equiv 2\theta-(\theta_++\theta_-)+\pi,
\label{small_sol}
\end{equation}
with $\theta_{\pm}=\tan^{-1}[u/(v\pm R)]$ and $x^2 = u^2 +v^2$,
which leads to a single monopole loop of radius $R$ in the $v$-plane 
(3-4 plane).
Note that $\beta_0$ has a jump form zero to $\pi$ at  $v=R$ 
in the $v$-plane, ($u=0$). This is where the magnetic monopole loop is located.

The above variational ansatz is in fact an exact 
solution to (\ref{beta_eq}) in the limit $R/\rho \rightarrow 0 $. 
Furthermore, in the limit  $R/\rho \rightarrow \infty $,
it corresponds to the gauge rotation that takes you from the singular gauge
to the non-singular gauge.
Thus one can say that the instanton in the non-singular gauge contains
a monopole loop of infinite radius, while the instanton in the singular
gauge contains a zero-size loop.

 We have found  numerically solutions that contain a monopole loop 
of arbitrary radius $R$ in the $v$-plane by setting  $\alpha = \phi - \psi $
and solving the remaining differential equation (\ref{beta_eq}). 
Finiteness of $G$ requires that $\beta(0,v)$ and $\beta(u,0)$ can only take on 
integer multiples of $\pi$. The loop of 
radius $R$ is introduced by enforcing the boundary conditions;  
at $\theta = 0$ ($v$-plane) $\beta$ has a jump
($0$ to $\pi$) at $v=R$ and at $\theta=\pi/2$ ($u$-plane) $\beta=\pi$. 
\begin{figure}[htb]
\vspace{9pt}
\psfig{file=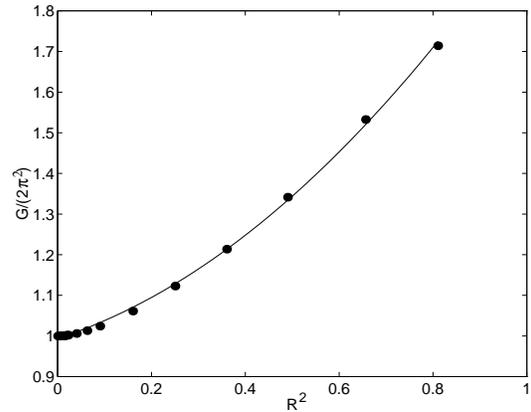,width=7cm}
\caption{G vs. R as computed by numerically solving the PDE. }
\label{fig:G-R}
\end{figure}
 The value of $G$ as a function of $R$ is plotted in Fig.\ref{fig:G-R}.
Note  that  $G$ is monotonically increasing with $R$. At small
$R/\rho$ it goes like $(R/\rho)^4log(R/\rho)$. Although $R=0$
corresponds to the true minimum, the
absence of a $R^2$-behavior corresponds to the presence of 
a ``zero-mode" so that infinitesimal-size  loops are easily produced.
Thus interactions or quantum fluctuations may
excite these loop modes. In fact we have done numerical experiments
which demonstrate that small deformations
in the instanton potential make finite loops favorable.

\section{4D Euclidean Lattice}
 
 In order to see if there are other configurations that might give
lower values to the MA functional  we did a lattice minimization of $G$,
approximated by
$$G = \frac{a^2}{2} \sum_{\mu,x}\{1-\frac{1}{2}
Tr( \Phi(x) U_\mu(x) \Phi(x+\mu)U^\dagger_\mu(x) )  \}.$$
Since we expect  at infinity that the functional drops like $1/x^4$ 
and that the gauge rotation is ${\bf 1}$, we can use a large enough 
volume with open boundary conditions to approximate $G$.
That will also help us put on the lattice a ``better'' instanton,
free of defects at the boundaries.

 Our minimization show that there is magnetic current loop formation 
on the lattice, the gauge rotation satisfies our ansatz; i.e.
$\alpha = \phi - \psi $  and $\beta$ only depends on $u$ and $v$,
but the radius $R$ does not scale with the instanton size $\rho$.
 This is in contradiction with
what was observed by Hart and Teper\cite{Hart}. Perhaps their use of
periodic boundary conditions causes the loop radius to scale
via interactions with the images of the instanton.
\begin{figure}[htb]
\vspace{9pt}
\psfig{file=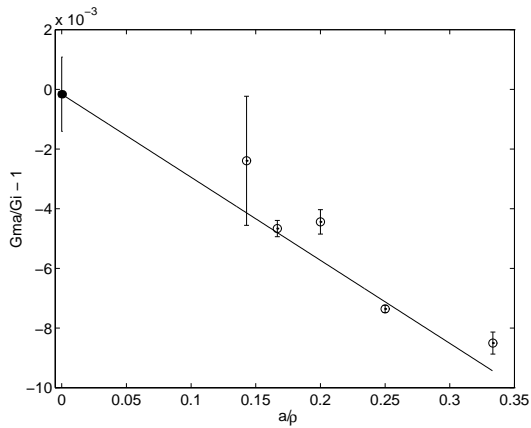,width=7cm}
\caption{Extrapolation to the continuum limit gives
 $ G_{ma}/G_{i} = ( 1.000 \pm 0.001) $. $G_i \equiv G[A^s_\mu]$.}
\label{fig:ext2}
\end{figure}

 If the loop is a lattice artifact, in the continuum limit it should
shrink to zero in physical units, and the MA functional should achieve
the value $4\pi^2 \rho^2$. We did finite size and volume analysis and
we were able to extrapolate to the continuum (infinite volume and zero
lattice spacing). We found that indeed in the continuum limit $G$ is
$4\pi^2 \rho^2$ within errors (Fig. \ref{fig:ext2}).

 We also studied the interacting case. Although this work is still in
progress, we find clear evidence that the instanton anti-instanton
(I-A) system does possess a monopole loop that survives the continuum
limit. There is a critical distance for which the individual loops
fuse into a single loop orbiting the I-A pair.  On the other hand, the
instanton-instanton system does not give rise to loops that survive
the continuum limit.

\section{Conclusions}

 The bottom line of our study is that the single instanton in the maximally
Abelian gauge possesses a monopole loop. We find that the global minimum
of $G$ for the isolated instanton is the singular gauge which is equivalent to
a zero sized monopole loop. The behavior of $G$ at small $R$ indicates
that quantum fluctuations or interactions with (anti)instantons probably
cause large loop formation. Our numerical study shows that
interactions with anti-instantons cause large loop formation.


\begin{thebibliography}{99}
\bibitem{Markum} H. Markum, W.Sakuler, S. Thurner, hep-lat/9510024
\bibitem{Schierh} V. Bornyakov, G. Schierholtz DESY 96-069, HLRZ 96-22,
                  hep-lat/9605019
\bibitem{Tanaka} H. Suganuma, A. Tanaka, S. Sasaki, O. Miyamura,
                 Nucl. Phys. B(Proc. Supp.), Proc. of Lattice 95.
\bibitem{Gubar} M.N. Chernodub and F.V. Gubarev, ITEP-95-34, hep-th/9506026
\bibitem{bto} R.C. Brower, K.N. Orginos and C-I Tan, BROWN-HET-1041
\bibitem{KSW} A. S. Kronfeld, G. Schierholz and Wiese, 
              Nucl. Phys. B293 (1987) 461.
\bibitem{Hart} A. Hart and M. Teper Oxford preprint No: OUTP-95-44-P,
              hep-lat/9511016 
\end{thebibliography}
\end{document}